\documentclass[twocolumn,amsmath,amssymb,floatfix]{revtex4}

\usepackage{latexsym,amsmath,amssymb,theorem,epsfig}
\usepackage{graphicx}

\usepackage{bm}
\usepackage{amsmath}

\def\lsim{\mathrel{\mathstrut\smash{\ooalign{\raise2.5pt\hbox{$<$}\cr\lower2.5pt\hbox{$\sim$}}}}}
\def\gsim{\mathrel{\mathstrut\smash{\ooalign{\raise2.5pt\hbox{$>$}\cr\lower2.5pt\hbox{$\sim$}}}}}

\def\myputfigure#1#2#3#4#5%
{\vskip#5pt\makebox[0pt]{\hskip#2in
\includegraphics[width=#3\textwidth]{#1}}\vskip#4pt\hfill}

\def\be{\begin{equation}}
\def\ee{\end{equation}}
\def\bea{\begin{eqnarray}}
\def\eea{\end{eqnarray}}

\def\({\left(}
\def\){\right)}

\begin{document}

\title{Flat 3-Brane with Tension in Cascading Gravity}

\author{Claudia de Rham$^{1,2}$, Justin Khoury$^{1,3}$, Andrew Tolley$^{1}$}

\affiliation{$^1$Perimeter Institute for Theoretical Physics,
31 Caroline St. N., Waterloo, ON, N2L 2Y5, Canada\\
$^{2}$Department of Physics \& Astronomy, McMaster University, Hamilton ON, L8S 4M1, Canada\\
$^{3}$Center for Particle Cosmology, University of Pennsylvania, Philadelphia, PA 19104--6395, USA}

\begin{abstract}
In the Cascading Gravity brane-world scenario, our 3-brane lies within a succession of lower-codimension branes, each with their own induced gravity term,
embedded into each other in a higher-dimensional space-time. In the 6+1-dimensional version of this scenario, we show that a 3-brane with tension remains
flat, at least for sufficiently small tension that the weak-field approximation is valid. The bulk solution is nowhere singular and remains in the perturbative regime everywhere.
\end{abstract}

\maketitle

An old idea to address the vexing problem of the cosmological constant is to confine the visible universe on a 3-brane in a higher-dimensional
space-time: vacuum energy on the brane curves the extra dimensions, but leaves the $4d$ geometry flat~\cite{rubshaposh}. While tantalizing, this proposal fails as soon as the extra dimensions are compactified; since $4d$ general relativity is recovered below the compactification scale, the theory unavoidably succumbs to Weinberg's no-go theorem~\cite{weinberg}. (An alternative strategy is to use compact extra dimensions to suppress radiative corrections to the cosmological constant~\cite{cliff}.)

The situation is drastically different, and far more promising, if the extra dimensions have infinite volume~\cite{dilute}. In this case, gravity is approximately $4d$ only
at short distances, thanks to an Einstein-Hilbert term on the brane, but becomes {\it higher-dimensional} in the infrared. In the DGP scenario~\cite{DGP} with one extra dimension, the gravitational force law on the brane scales as the usual $1/r^2$ at short distances, but asymptotes to $1/r^3$ at large distances. Gravity therefore behaves as a high-pass filter~\cite{filter}. This weakening of gravity suggests that vacuum energy, by virtue of being the longest-wavelength source, only {\it appears} small because it is {\it degravitated}~\cite{filter,degrav}.

The degravitation phenomenon is not realized in the original DGP model because the weakening of the force law is too shallow in the infrared~\cite{degrav}. This motivates exploring higher-codimension branes, {\it i.e.}, a higher-dimensional bulk.
Realizing these higher-codimension scenarios has proven difficult. To begin with, the simplest
constructions are plagued by ghost instabilities~\cite{dubov,Gabadadze:2003ck}. Secondly, the $4d$ propagator is divergent
and must be regularized~\cite{cod2eft}. Furthermore, for a static bulk, the geometry for codimension $N> 2$ has
a naked singularity at finite distance from the brane, for arbitrarily small tension~\cite{dilute}.
(Allowing the brane to inflate gives a Hubble rate on the brane {\it inversely}
proportional to the brane tension for codimension $N > 2$~\cite{dilute}.)

Recently, it was argued that these pathologies are resolved
by embedding our 3-brane within a succession of higher-dimensional
branes, each with their own induced gravity
term~\cite{cascade1,cascade2,claudiareview}. We refer to this
framework as Cascading Gravity. The induced graviton kinetic term
acts as a regulator for the 3-brane propagator~\cite{cascade1,cascade2}.
In the case $N=2$ studied in~\cite{cascade1}, consisting of a 3-brane embedded in a 4-brane
within a 5+1-dimensional bulk, the ghost is cured by including a
sufficiently large tension on the (flat)
3-brane~\cite{cascade1,upcoming}.
Alternatively, the ghost is also cured when considering a
higher-dimensional Einstein-Hilbert term localized on the brane,
\cite{Gabadadze:2003ck,cascade2}.

\begin{figure} 
   \centering
   \includegraphics[width=0.4\textwidth]{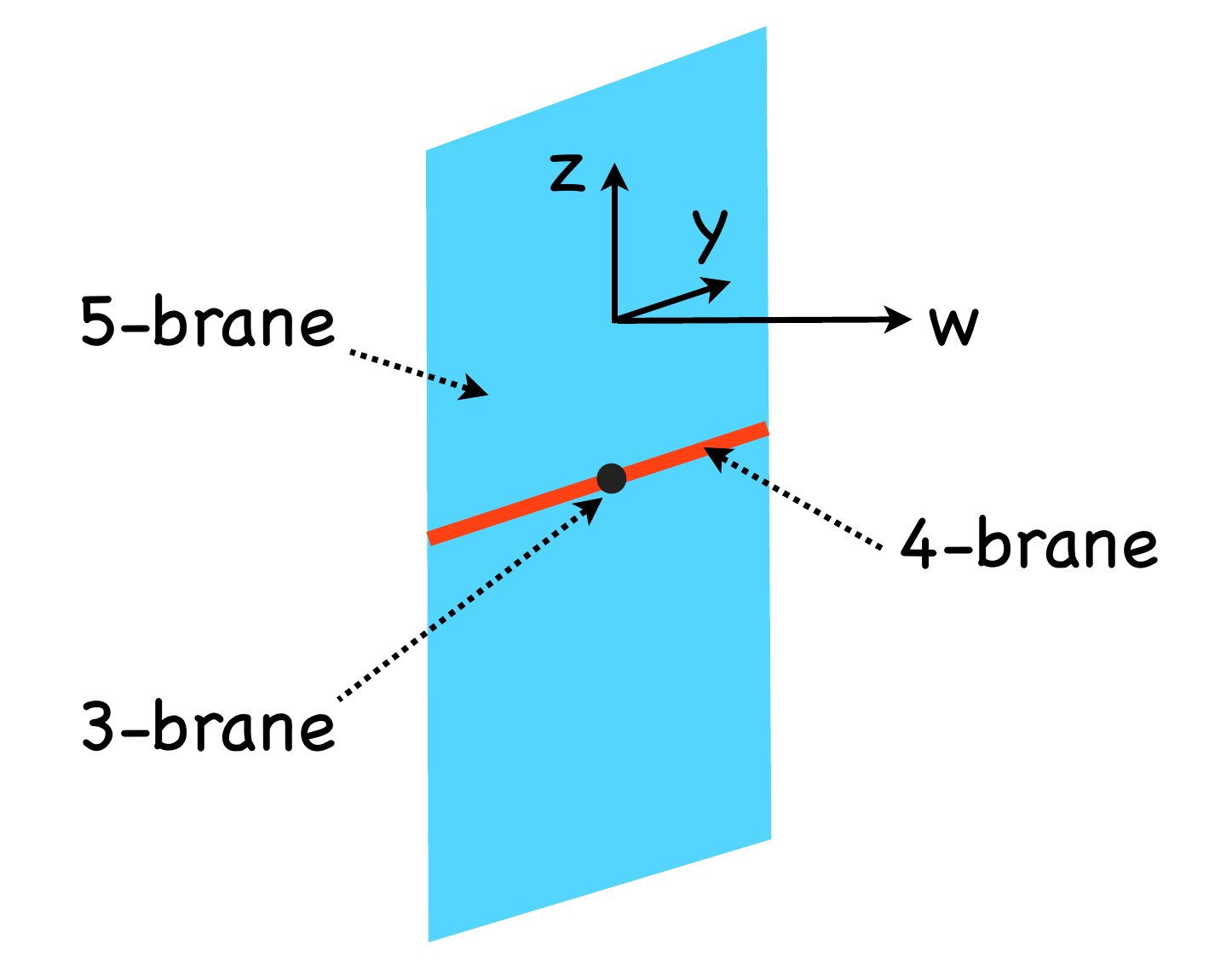}
   \caption{Sketch of the codimension-3 cascading gravity set-up.}
   \label{cod3sketch}
\end{figure}

Already with $N=2$ the solution exhibits degravitation: a 3-brane with tension creates a deficit angle in the bulk
while remaining flat~\cite{upcoming}. We stress that this self-tuning mechanism crucially relies on the
extra dimensions having infinite volume: if the dimensions were compact, the brane tension would have to be tuned against
other branes and/or bulk fluxes~\cite{sumrule}.

Since the deficit angle must be less than $2\pi$, the tension allowed by the solutions considered in \cite{cascade1,upcoming} is bounded by $M_6^4$, where the $6d$ Planck mass $M_6$ is itself
constrained to be at most $\sim$meV. Given its geometrical nature, this bound is most likely an artifact of the codimension-2 case and is expected to be absent
in higher-codimension.

Motivated by these considerations, in this Letter we explore Cascading Gravity with $N=3$, consisting of a 3-brane living on a 4-brane, itself embedded in a 5-brane, together in a 6+1-dimensional bulk, as sketched in Fig.~\ref{cod3sketch}. Including tension on the 3-brane, we derive a solution for which $i)$ the bulk metric is non-singular everywhere (except, of course, for a delta-function in curvature at the 3-brane location) and asymptotically flat; and $ii)$ the induced 3-brane geometry is exactly flat.

Since the metric depends on 3 spatial coordinates, to proceed analytically we restrict ourselves to the weak-field approximation, corresponding to sufficiently
small tension. For consistency, we check that our solution remains perturbative everywhere. We are currently working on numerically extending these solutions
to the non-linear regime of large tension.

Unlike the case of a pure codimension-3 DGP brane in 6+1 dimensions, where the static bulk geometry has a naked singularity for arbitrarily small tension~\cite{dilute}, here the bulk metric is completely smooth. This traces back to the cascading mechanism of regulating the propagator: the presence of parent branes removes the power-law divergence in the $4d$ propagator.

As illustrated in Fig.~\ref{cod3sketch}, the 3 extra spatial dimensions are denoted by $y,z$ and $w$, with the codimension-1
brane located at $w=0$, the codimension-2 brane at $z=w=0$, and the codimension-3 brane at $y=z=w=0$.
Indices in $7d$ are denoted by $A,B,\ldots$, indices in $6d$ by $a,b,\ldots$,
in $5d$ by $\alpha,\beta,\ldots$, and finally in $4d$ by $\mu,\nu,\ldots$

\noindent {\bf I. Scalar Green's Functions}: In solving for the metric perturbations, it is useful to first consider the scalar Green's functions, determined from the action
\bea
S&=&\frac{1}{2} \int {\rm d}^7x \; \Psi \left[M_7^5 \Box_7 + \delta(w)M_6^4\Box_6 \right. \\
&+& \left. \delta^2(z,w)M_5^3 \square_5 + \delta^3(y,z,w)M_4^2 \Box_4 \right]\Psi\,, \nonumber
\eea
where $M_d$ denotes the ``Planck" mass in $d$ dimensions. The model has three cross-over scales,
\be
m_5 = \frac{M_5^3}{M_4^2}\;;\;\;\; m_6 = \frac{M_6^4}{M_5^3}\;; \;\;\; {\rm and} \;\;\; m_7 = \frac{M_7^5}{M_6^4}\,,
\ee
marking, respectively, the transition scale from $4d$ to $5d$, from $5d$ to $6d$, and finally from $6d$ to $7d$.

In the absence of the $5d$ and $4d$ kinetic terms, the propagator on the codimension-1 brane is
of the DGP form~\cite{DGP}
\be
G_{6}^{(0)}(z-z') = \frac{1}{M_6^4}\int \frac{{\rm d}\omega}{2\pi}\frac{e^{i\omega (z-z')}}{\omega^2+q^2+m_7\sqrt{q^2+\omega^2}}\,,
\label{G6}
\ee
where $q^\alpha$ is the $5d$ momentum, and $\omega$ is the momentum associated with the $z$ coordinate.
The exact $6d$ propagator is then obtained by treating the $5d$ kinetic term
as a perturbation localized at $z=0$:
\begin{align}
&G_{6}(z,z')= G_{6}^{(0)}(z-z') - M_5^3 G_{6}^{(0)}(z) q^2 G_{6}^{(0)}(-z') \cr &~~~~~~~~~+ M_5^6
G_{6}^{(0)}(z) q^4 G_{6}^{0)}(0) G_6^{(0)}(-z')+\dots \cr
&=G_6^{(0)}(z-z')- \frac {G_6^{(0)}(z)M_5^3 q^2G_6^{(0)}(-z')} {1 + M_5^3 q^2 G_6^{(0)}(0)}
\,.
\end{align}
In particular the induced propagator on the codimension-2 brane is determined
in terms of the integral of the higher dimensional Green's function:
\be
G_{5}^{(0)}(q^2)=G_6(0,0)= \frac {1} {M_5^3}\frac{1}{q^2+g(q^2)}\,;
\label{G5}
\ee
\be
g(q^2) \equiv \frac{1}{M_5^3G^{(0)}_6(0)} = \frac{\pi m_6}{2}\frac{\sqrt{m_7^2-q^2}}{\tanh^{-1}
\,\left(\sqrt{\frac{m_7-|q|}{m_7+|q|}}\right)}\,.
\label{gp}
\ee
(For $|q|>m_7$ we assume analytic continuation from hyperbolic tangent to its trigonometric
counterpart.)

Remarkably, the codimension-1 kinetic term makes the $5d$ propagator finite,
thereby regulating the logarithmic divergence characteristic of pure codimension-2
branes. Indeed, $G_5^{(0)} \rightarrow M_7^{-5}\log(m_7q)$ as $M_6\rightarrow 0$,
and thus $M_6$ plays the role of a physical cut-off. As another check, note that
in the limit $m_7\rightarrow 0$ in which the bulk decouples, we recover the usual DGP
result: $G_5^{(0)} \sim 1/(q^2 + m_6q)$.

It is straightforward to repeat the same steps to derive the induced $4d$ propagator on
the codimension-3 brane.

\noindent {\bf II. Cascading Gravity}: We now proceed to the gravitational case. The $7d$ Einstein equations are given by
\bea
\nonumber
M_7^5G_{AB}^{(7)} &=& -\delta(w)\left\{\delta_A^{\;a}\delta_B^{\;b}M_6^4G_{ab}^{(6)} + \delta(z)\delta_A^{\;\alpha}\delta_B^{\;\beta}M_5^3G_{\alpha\beta}^{(5)}\right. \\
&+& \left. \delta(z)\delta(y)\delta_A^{\;\mu}\delta_B^{\;\nu}\left[M_4^2G_{\mu\nu}^{(4)} + \Lambda g_{\mu\nu}\right]\right\}\,.
\label{ein}
\eea
The effective source therefore consists of induced gravity terms on each of the branes, as well as tension $\Lambda$ on the codimension-3 brane. 

In the weak-field approximation, the $7d$ line element can be written as ${\rm d}s^2 = (\eta_{AB} +h_{AB}){\rm d}x^A{\rm d}x^B$. As
shown in Appendix I, there is enough symmetry and gauge freedom to simplify the metric to the form
\bea
\nonumber
{\rm d}s^2 &=& \left(1+\Phi(y,z,w)\right)\left({\rm d}w^2 + {\rm d}z^2 + {\rm d}y^2\right) \\
\nonumber
&- & \frac{\Theta(w)}{2m_7}\partial_\alpha\Phi_0(y,z){\rm d}x^\alpha{\rm d}w \\
&+& \left(1-\frac{\Phi(y,z,w)}{4}\right)\eta_{\mu\nu}{\rm d}x^\mu {\rm d}x^\nu\,,
\label{nonsing}
\eea
where $\Phi_0(y,z) \equiv \Phi(y,z,w=0)$ is the induced profile on the codimension-1 brane. Here $\Theta(w)$ is the theta function: $\Theta(w) = +1$ for $w>0$, and $-1$
for $w<0$.

\begin{figure} 
   \centering
   \includegraphics[trim = 10mm 0mm 30mm 10mm, width=0.4\textwidth]{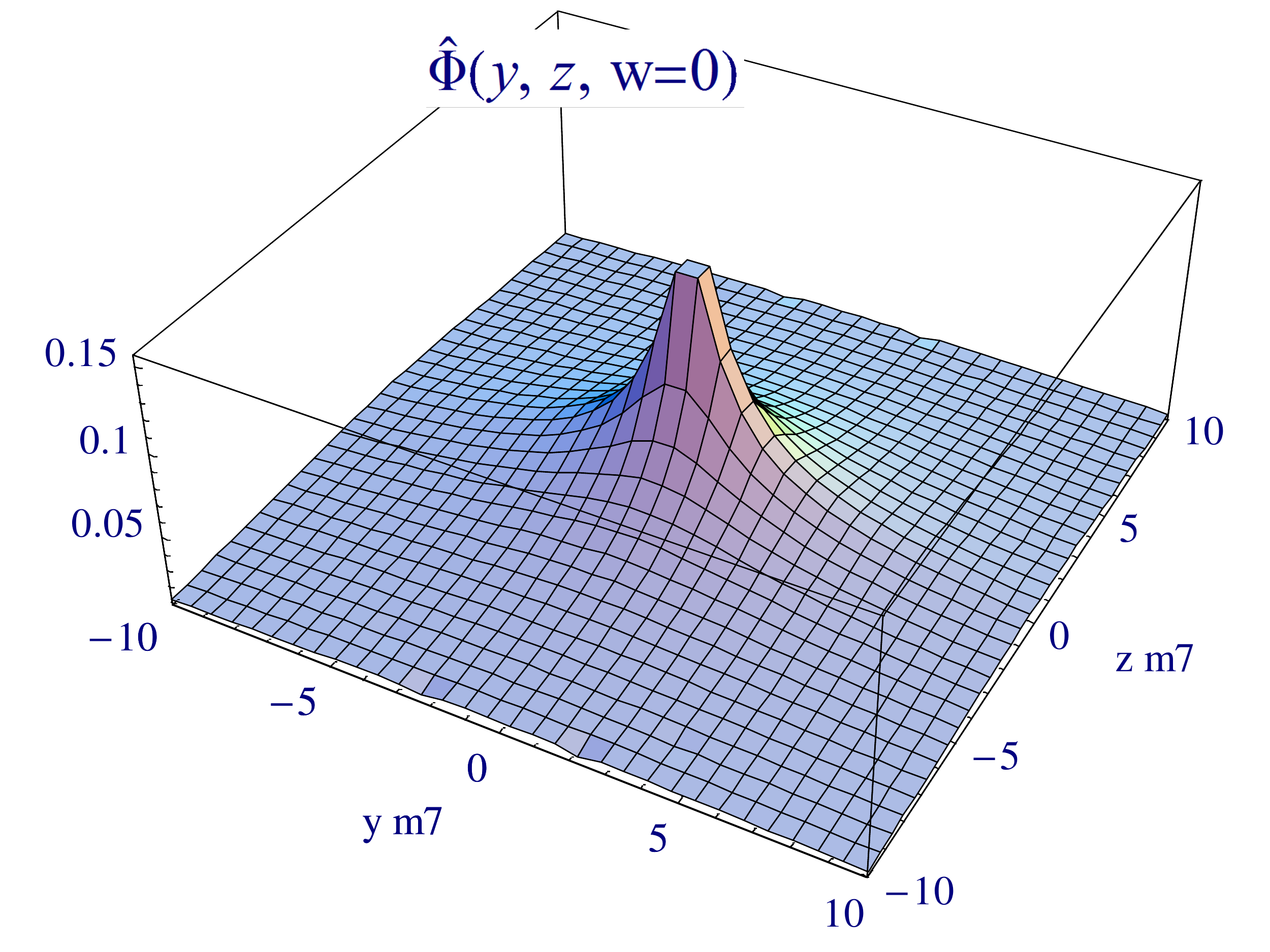}
   \includegraphics[trim = 10mm 10mm 30mm 0mm, width=0.4\textwidth]{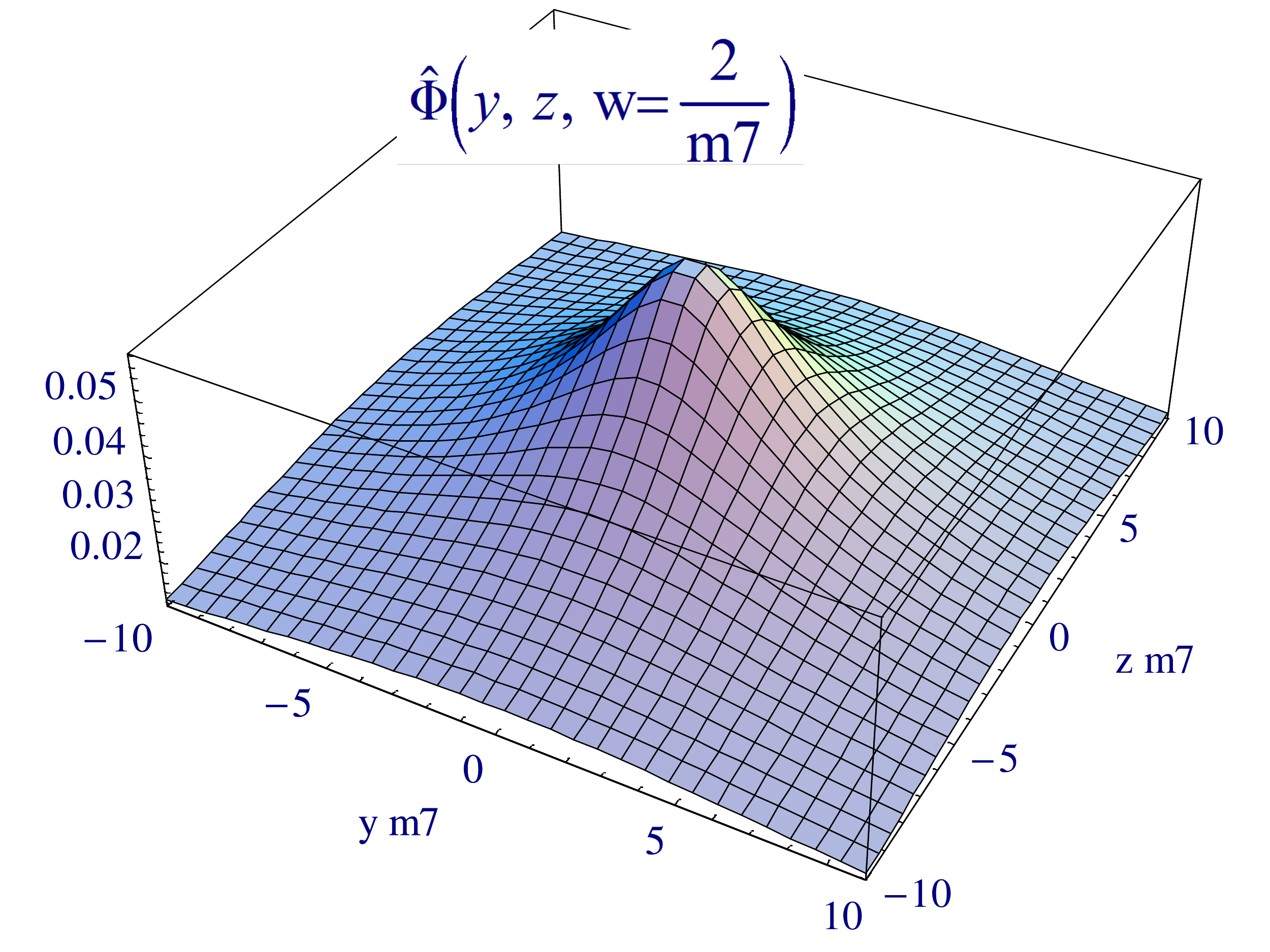}
   \caption{Plot of the solution for the metric potential $\hat \Phi(y,z,w)$ for $w=0$ and $w=2m_7^{-1}$
   in the case where $m_6=m_7$.}
   \label{numsol}
\end{figure}

Substituting this ansatz into Einstein's equations~(\ref{ein}), we find that $\Phi$ satisfies
\bea
\left(\Box_7 +\frac{\delta(w)}{m_7}\Box_6 -\frac{3}{5}\frac{\delta^2(z,w)}{m_7m_6}\Box_5\right)\Phi = \frac{8}{5}\frac{\delta^3(y,z,w)}{M_7^5}\Lambda\,. \ \
\label{fullPhi}
\eea
This equation is of the cascading form~\cite{cascade2}, as reviewed above.
The asymptotically flat bulk solution is given by
\be
\Phi(y,z,w) = e^{-|w|\sqrt{-\Box_6}}\Phi_0(y,z)\,,
\label{bulksol}
\ee
where the induced profile $\Phi_0(y,z)$ satisfies
\be
\left(\Box_6 - m_7\sqrt{-\Box_6} -\frac{3}{5}\frac{\delta(z)}{m_6}\Box_5\right)\Phi_0 = \frac{8}{5}\frac{\delta^2(y,z)}{M_6^4}\Lambda\,.
\label{cod1eom}
\ee
To solve~(\ref{cod1eom}), we Fourier transform to momentum space and use the $6d$ and $5d$ Green's functions
given respectively by~(\ref{G6}) and~(\ref{G5}). The result is
\be
\Phi_0(y,z) = \int \frac{{\rm d}q_y{\rm d}\omega}{(2\pi)^2} \frac{e^{i\omega z}e^{iq_yy} g(q_y) \phi(q_y)}{\omega^2+q_y^2+m_7\sqrt{\omega^2+q_y^2}}\,,
\label{cod1sol}
\ee
where the Fourier transform of the codimension-2 profile, $\phi(q_y) = \int {\rm d}y e^{-iq_y y}\Phi_0(z=0,y)$, satisfies
\be
\left(\frac{3}{5}q_y^2 - g\left(q_y^2\right)\right)\phi(q_y) = \frac{8}{5M_5^3}\Lambda\,.
\label{phiq}
\ee
The solution to~(\ref{phiq}) can be expressed as the sum of a principal part $\mathcal P$ and two homogeneous modes:
\be
\phi(q_y)=\frac{8\Lambda}{5M_5^3}\,\mathcal{P}\left[\frac{1}{\frac{3}{5}q_y^2 - g\left(q_y^2\right)}\right]
+\sum_{\sigma=\pm}C_\sigma\delta(q_y-\sigma q_0)\notag\,,
\ee
where $\frac 35 q_0^2=g(q_0^2)$. Requiring the field $\Phi_0$ to be real imposes $C_+=C_-\equiv C$, while requiring $\Phi_0$ to fall as $y\to 0$ sets $C=0$.  
Using the resulting expression for $\phi(q_y)$ into~(\ref{cod1sol}) and then into~(\ref{bulksol}), we obtain the
final expression for the scalar potential $\Phi(y,z,w) = \frac{8\Lambda}{5M_6^4} \hat
\Phi(y,z,w)$:
\bea
\hat \Phi \hspace{-2pt}= \hspace{-4pt}\int \frac{{\rm d}\omega {\rm d}q_y}{(2\pi)^2}  \frac{e^{-|w|\sqrt{\omega^2+q_y^2}}e^{i\omega z}e^{iq_yy}}{\omega^2+q_y^2+m_7\sqrt{\omega^2 + q_y^2}}
 \mathcal{P}\hspace{-3pt}\left[\frac{g(q_y)}{\frac{3}{5}q_y^2- g(q_y)}\right]\,
\eea
This is our main result. Thanks to the cascading mechanism, which has regularized all potential divergences, {\it this solution is finite everywhere}.
Figure~\ref{numsol} shows that $\hat \Phi(y,z,w)$ is smooth everywhere and decreases with $w$.

As it stands, however, our framework has a ghost~\cite{dubov,Gabadadze:2003ck}, as indicated by the poles at $q_y=\pm q_0$. 
There are two ways to resolve this issue. One can introduce sufficiently large tension on both the codimension-2 and
-3 branes~\cite{cascade1}: to remove the ghost, the codimension-2 tension should be $\;\gsim\; M_5^3 m_7^2$, whereas
the corresponding bound on the codimension-3 tension is yet to be determined.  

Alternatively, one can regularize codimension-2 and -3 branes and include $6d$ Einstein-Hilbert term localized
on these objects~\cite{Gabadadze:2003ck,cascade2}. Following this route, we demonstrate in Appendix II that
the poles do disappear, and that the profile for $\Phi(y,z,w)$ is qualitatively unchanged. 

\noindent {\bf III. Discussion}: In this Letter we have shown that a 3-brane with tension remains flat in the 6+1-dimensional cascading gravity
framework. In the weak-field approximation,  we have obtained a bulk solution which is nowhere singular and remains perturbative everywhere.

These properties crucially depend on the existence of parent branes with finite Planck masses. Indeed, our solution goes
outside the perturbative regime and acquires divergences in the limit $M_5, M_6\rightarrow 0$, consistent with~\cite{dilute}.

We are currently extending our solutions to the non-linear regime through numerical analysis. For now, we view the present
results as a tantalizing first step towards realizing the idea of Rubakov and Shaposhnikov.

{\it Acknowledgments:} We thank G.~Dvali and G.~Gabadadze for helpful discussions. This work was supported in part by NSERC and Ontario's MRI.

\noindent {\bf Appendix I}: We show that the weak-field metric can be brought to the form~(\ref{nonsing}) by symmetry and gauge freedom. 
In de Donder gauge, $\partial_A h^A_{\; B} = \frac{1}{2}\partial_B h^C_{\; C}$,~(\ref{ein}) reduces to
\be
\nonumber
-\frac{M_7^5}{2}\Box_7\left(h_{AB}-\frac{1}{2}\eta_{AB}h^C_{\; C}\right) = \delta(w)\left(T_{ab}^{(6)} -M_6^4G_{ab}^{(6)}\right)\,,
\label{7d}
\ee
where the effective stress-energy on the codimension-1 brane, $T_{ab}^{(6)}$, includes contributions from the $5d$ and $6d$ induced gravity terms.
Since there is no stress energy along the $(a,w)$ and $(w,w)$ directions, the corresponding
equations are consistently satisfied by setting $h_{aw} = 0$ and $h_{ww} =h^c_{\; c}$
where $h^c_{\; c}$ is the $6d$ trace. It follows that the induced gauge choice in $6d$ is given by $\partial_a h^a_b = \partial_b h^c_{\;c}$, hence
the $(a,b)$ components of~(\ref{7d}) reduce to
\bea
\nonumber
-\frac{M_7^5}{2}\Box_7\left(h_{ab}-\eta_{ab}h^c_{\; c}\right) &=& \delta(w) \frac{M_6^4}{2}\left(\Box_6h_{ab}-\partial_a\partial_bh^c_{\; c}\right) \\
&+& \delta(w)T_{ab}^{(6)}\,.
\label{6dein}
\eea

To proceed further, it is convenient to decompose $h_{ab}$ into its trace and transverse-traceless (TT) parts:
\be
h_{ab} = h_{ab}^{6d{\rm TT}} +  \frac{\partial_a\partial_b}{\Box_6}h^c_{\; c}\,.
\label{decomp1}
\ee
From~(\ref{6dein}), the $6d$TT components satisfy
\bea
\nonumber
&  & - \frac{M_7^5}{2}\left(\Box_7 + \frac{\delta(w)}{m_7}\Box_6\right)h_{ab}^{6d{\rm TT}} = \\
& & \;\;\;\;\;\; \delta(w)  \left(T_{ab}^{(6)}-\frac{1}{5}\eta_{ab}T^{(6)}+\frac{1}{5}\frac{\partial_a\partial_b}{\Box_6}T^{(6)}\right)\,.
\label{6dTT}
\eea

The symmetries of the problem allow a simple expression for the $5d$ components of the $6d$TT part:
\be
h_{\alpha\beta}^{6d{\rm TT}} = - \frac{1}{4}\Phi\eta_{\alpha\beta} - \left(\frac{\Box_5}{\Box_6}-\frac{5}{4}\right)\frac{\partial_\alpha\partial_\beta}{\Box_5}\Phi\,.
\label{Phidecomp}
\ee
This follows from setting $h_{\alpha\beta}^{5d{\rm TT}} = 0$, which is consistent with the equations of motion for the case of interest. Substituting into~(\ref{6dTT}), and using $T^{(5)}_{\alpha\beta} = -\delta^\mu_{\; \alpha}\delta^\nu_{\; \beta}\Lambda\eta_{\mu\nu}\delta(y)$, the resulting equation of motion for $\Phi$ agrees with~(\ref{fullPhi}).

We can now be explicit about the form of the various metric components. Combining~(\ref{decomp1}) and~(\ref{Phidecomp}),
we get:
\be
h_{\alpha\beta} = -\frac{1}{4}\Phi\eta_{\alpha\beta} - \left(\frac{\Box_5}{\Box_6} - \frac{5}{4}\right)\frac{\partial_\alpha\partial_\beta}{\Box_5}\Phi + \frac{\partial_\alpha\partial_\beta}{\Box_6}h^c_{\; c}\,.\
\ee
And since everything is independent of $x^\mu$, we get $ h_{y\mu} = 0$
and $h_{\mu\nu} =-\frac{1}{4}\Phi\eta_{\mu\nu}$. Similarly, from~(\ref{decomp1}) we obtain
\bea
\nonumber
h_{yz} &=& \frac{\partial_y\partial_z}{\Box_6}\left(h^c_{\; c} - \Phi\right)\;; \;\; h_{zz} = \frac{\partial^2_z}{\Box_6}\left(h^c_{\; c} - \Phi\right)  + \Phi\;; \\
h_{yy} &=& \frac{\partial^2_y}{\Box_6}\left(h^c_{\; c} - \Phi\right)  + \Phi \,.
\eea
The is equivalent to (\ref{nonsing}) after a small diffeomorphism.

\noindent {\bf Appendix II}: One way to cure the ghost of higher-codimension DGP models~\cite{dubov,Gabadadze:2003ck} is to consider a
higher-dimensional Einstein-Hilbert term localized on the regularized brane~\cite{Gabadadze:2003ck,cascade2}.
Following this prescription, we will show that the solution remains finite everywhere. 

When adding a $6d$ Einstein-Hilbert term on the regularized 4-brane, 
on the top of the usual $5d$ Einstein-Hilbert term of the form `$\Box_5 h_{\alpha \beta}$'
we must consider excitations of transverse modes along the extra dimensions
as well as the higher-dimensional mode $h_{zz}$. In the thin-brane limit, however, the excitations along the extra dimension become very massive,
so that any term containing $z$-derivatives can be neglected. Meanwhile, $h_{zz}$ survives in the limit. See~\cite{cascade2} for details.

In 7$d$ de Donder gauge, the Einstein equations are the same
as in~\eqref{7d}. Setting $h_{aw} = 0$ and $h_{ww} =h^c_{\; c}$, we have
\bea
-\frac{M_7^5}{2}\(\hspace{-2pt}\Box_7+\frac{\delta(w)}{m_7}\Box_6\hspace{-2pt}\)\hspace{-2pt}h_{ab}=\delta(w)\hspace{-2pt}\(\hspace{-2pt}T^{(6)}_{ab}-\frac 15 T^{(6)}\eta_{ab}\hspace{-2pt}\)
\eea
with $T^{(6)}_{z\alpha}=0$, $T^{(6)}_{zz}= M_5^3\delta(z)R_5/2$, and
\bea
\notag &&\hspace{-10pt}T^{(6)}_{\alpha \beta}=-M_5^3\delta(z)\hspace{-3pt}\left[G_{\alpha \beta}^{(5)}
+\frac 12 \(\Box_5 h_{zz}\eta_{\alpha\beta}-\partial_\alpha\partial_\beta h_{zz}\)
\right]\ \\
&&\hspace{20pt}-\delta(z)\delta(y)\Lambda
\eta_{\mu\nu}\delta^\mu_\alpha\delta^\nu_\beta\,.
\eea
Using this in the $6d$ part of the Einstein equations, we get 
$h_{zz}=-\psi$, $\Box_5h_{yy}=-4\Box_5\psi+\partial_y^2h^\alpha_{\, \alpha}$,
$h_{\mu y}=0$ and $\Box_5h_{\mu \nu}=\Box_5\psi\eta_{\mu \nu}+\partial_\mu\partial_\nu h^\alpha_{\, \alpha}$,
with
\bea
\left[ \Box_7+\frac{\delta(w)}{m_7} \Box_6+\frac{\delta^{(2)}(w,z)}{m_7 m_6} \Box_5\right]\psi=\frac{2}{5} \frac{\delta^{(3)}(w,z,y)}{M_7^5}\Lambda\,. \ \
\label{psieom}
\eea
We notice that the kinetic term for $\psi$ is now everywhere positive, signaling that the ghost has been cured.
Equation~(\ref{psieom}) is similar to~\eqref{fullPhi} for $\Phi$, except for a redefinition of $m_6$ and $M_7$.
The profile for $\psi(y,z,w) = -\frac{2\Lambda}{5M_6^4} \hat \Psi(y,z,w)$,
\bea
\hat \Psi =\int \frac{{\rm d}\omega {\rm d}q_y}{(2\pi)^2}  \frac{e^{-|w|\sqrt{\omega^2+q_y^2}}e^{i\omega z}e^{iq_yy}}{\omega^2+q_y^2+m_7\sqrt{\omega^2 + q_y^2}}
\frac{g(q_y)}{q_y^2+ g(q_y)}\,,
\eea
is similar to that of $\hat\Phi$, and, in particular, is free of divergences.
The static solution for a codimension-3 brane with tension remains therefore well-defined,
at least in the weak field approximation, in a ghost-free set-up.

\end{document}